\documentclass[aps,prl,reprint,superscriptaddress]{revtex4-1}
\usepackage{cases}
\usepackage{xcolor}
\usepackage{braket}
\usepackage{graphicx}
\usepackage{longtable}

\begin{document}


\title{Laser cooling in a chip-scale platform}

\author{J.~P.~McGilligan}
\email{jpmcgilligan91@gmail.com}
\affiliation{University of Colorado, Department of Physics, Boulder, Colorado, 80309, USA}
\affiliation{National Institute of Standards and Technology, Boulder Colorado, 80305, USA}
\author{K.~R.~Moore}
\affiliation{National Institute of Standards and Technology, Boulder Colorado, 80305, USA}
\author{A. Dellis}
\affiliation{University of Colorado, Department of Physics, Boulder, Colorado, 80309, USA}
\affiliation{National Institute of Standards and Technology, Boulder Colorado, 80305, USA}
\author{G. D. Martinez}
\affiliation{University of Colorado, Department of Physics, Boulder, Colorado, 80309, USA}
\affiliation{National Institute of Standards and Technology, Boulder Colorado, 80305, USA}
\author{E.~de Clercq}
\affiliation{LNE-SYRTE, Observatoire de Paris, Universit\'e PSL, CNRS, Sorbonne Universit\'e, Paris, France.}
\author{P.~F.~Griffin}
\affiliation{Department of Physics, SUPA, University of Strathclyde, Glasgow, G4 0NG, UK}
\author{A.~S.~Arnold}
\affiliation{Department of Physics, SUPA, University of Strathclyde, Glasgow, G4 0NG, UK}
\author{E.~Riis}
\affiliation{Department of Physics, SUPA, University of Strathclyde, Glasgow, G4 0NG, UK}
\author{R.~Boudot}
\affiliation{FEMTO-ST, CNRS, 26 chemin de l`Epitaphe, 25030 Besan\c con, France}
\affiliation{National Institute of Standards and Technology, Boulder Colorado, 80305, USA}
\author{J.~Kitching}
\affiliation{National Institute of Standards and Technology, Boulder Colorado, 80305, USA}



\date{\today}

\begin{abstract}
Chip-scale atomic devices built around micro-fabricated alkali vapor cells are at the forefront of compact metrology and atomic sensors. We demonstrate a micro-fabricated vapor cell that is actively-pumped to ultra-high-vacuum (UHV) to achieve laser cooling. A grating magneto-optical trap (GMOT) is incorporated with the 4~mm-thick Si/glass vacuum cell to demonstrate the feasibility of a fully-miniaturized laser cooling platform. A two-step optical excitation process in rubidium is used to overcome surface-scatter limitations to the GMOT imaging. The unambiguous miniaturization and form-customizability made available with micro-fabricated UHV cells provide a promising platform for future compact cold-atom sensors.
\end{abstract}

\pacs{}

\maketitle

\section{Introduction}
Laser-cooled atomic samples have led to significant advances in precision metrology as a result of  increased measurement time and reduced environmental interactions compared to room temperature ensembles. The magneto-optical trap (MOT) is commonly used as a source of cold atoms in state-of-the-art atomic clocks and interferometers \cite{Hinkley2013,Kovachy2015}. Although there has been important progress toward portable systems and instruments based on laser-cooled atoms \cite{Laurent,Bongs} for use on mobile platforms such as vehicles, aircraft and ships, these systems still typically occupy volumes of $>$100~L and consume many watts of power to operate.

Instruments based on room-temperature atomic ensembles \cite{Knappe2004,Schwindt2004} have been successfully miniaturized using micro-electro-mechanical-systems (MEMS) alkali vapor cells \cite{Liew2004} containing buffer gases to reduce relaxation due to wall collisions. However, buffer-gas mixtures require critical temperature stabilization to reduce sensitivity to pressure shifts and introduce large frequency offsets that prohibit accurate realization of the SI second \cite{Kitching2018}.

Further miniaturization to sub-liter volumes is also underway. The generation of cold-atoms have been demonstrated using a single laser beam reflected from a pyramidal \cite{Trupte2006,arnold1} or cone-shaped structures \cite{Lee96,Shah2011}, and novel cell geometries are being explored to reduce size \cite{sebby2016,selim}. On-chip magnetic confinement of laser-cooled atoms \cite{Reichel1999,Folman} also offers unique opportunities for compact instrumentation \cite{ramirez}, but such traps are still typically loaded from a liter-volume vacuum apparatus.

Recent advancements in the miniaturization of cold atom packages have focused on the micro-fabrication of optical elements \cite{McGilligan2017}, alkali vapor density regulators \cite{Kang2019,mcgilliganPRApplied} and low-power coils \cite{Saint2018} to facilitate a compact cold-atom device. However, the miniaturization of UHV vacuum packages \cite{siliconbulk} remains limited to bulk machining of chamber materials, which lacks the scalability made possible with micro-fabrication. 

Recent work has showed that anodic bonding of silicon frames and aluminosilicate glass (ASG) can reduce the effects of He permeation to a level that would in principle allow a MOT to persist for six months without active pumping \cite{Dellis2016}. The use of passive pumping in MEMS cells could mitigate the scalability and power consumption limitations placed by currently available active pumping mechanisms.

In this paper, we demonstrate laser cooling of $^{85}$Rb in a glass-silicon-glass UHV cell, with a traditional 6-beam MOT connected to an ion pump for active pumping. The apparatus is miniaturized further by using a micro-fabricated grating chip in conjunction with a micro-fabricated cell to realize a compact 4-beam MOT with only a single input beam. The association of these two technologies constitutes a promising solution for the development of miniaturized MOTs, compatible with lithographic fabrication techniques. 

However, in such a set-up imaging of the MOT at $780\,$nm is made difficult due to background light, scattered from the grating chip surface and Si cell walls. To overcome this issue, a two-step excitation to the 5$D_{5/2}$ level in rubidium is used to generate fluorescence from the laser-cooled atoms at $420\,$nm, which with appropriate filtering, overcomes surface scatter and enables imaging of the cold atoms even with the low atom numbers that results from the small beam overlap volume \cite{Ohadi2009,Sheludko2008}. With a total cell volume of $4\,$cm$^3$, this amalgamation of micro-fabricated components is a promising platform for future compact cold-atom sensors and instruments.

\section{Experimental set-up}

\begin{figure}[t]
\centering
\includegraphics[width=\columnwidth]{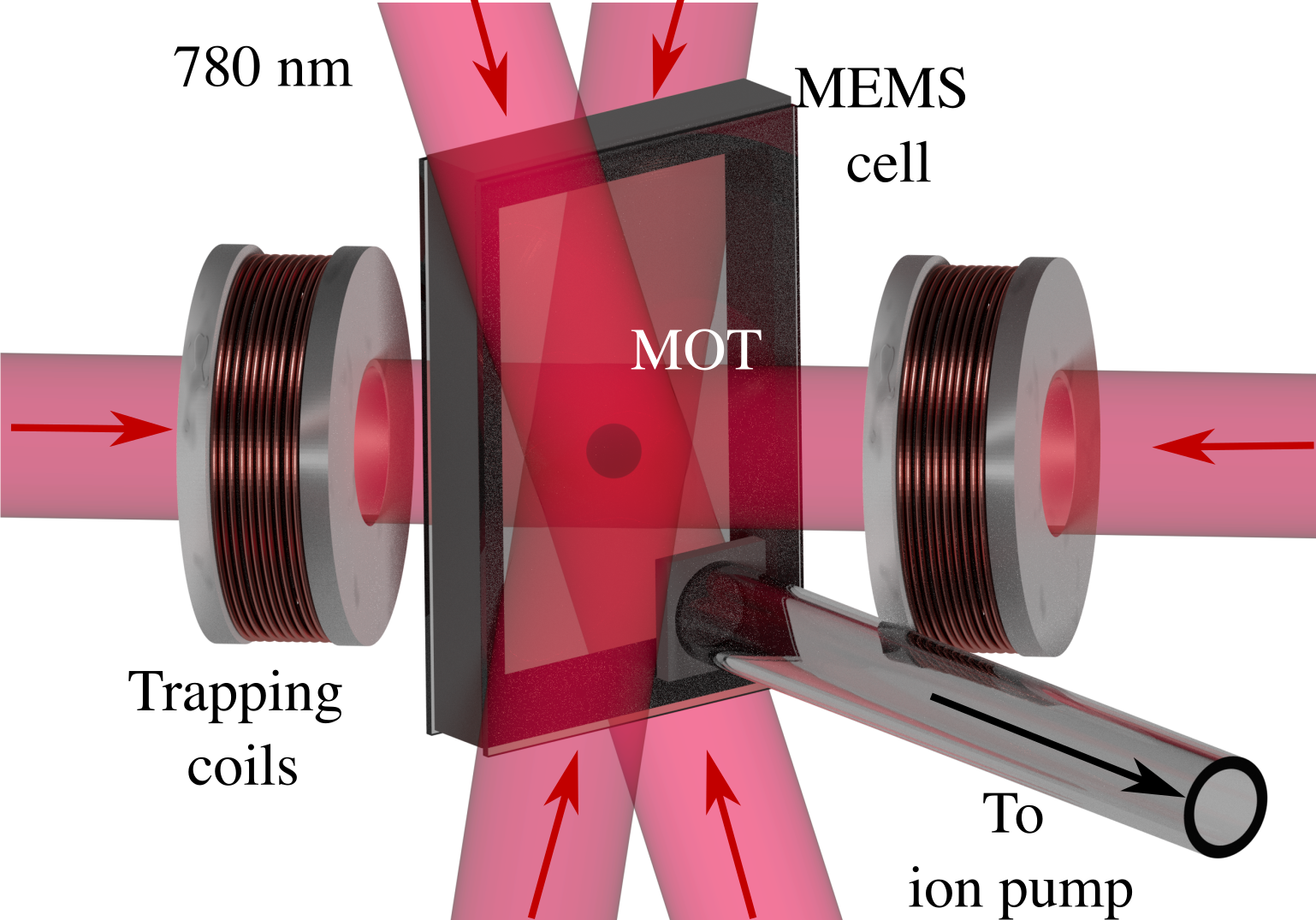}
\caption{ MEMS vacuum cell. An etched silicon frame is anodically bonded at upper and lower surfaces to glass wafers. The top glass wafer is drilled and bonded to a Si washer and glass tube. The MEMS cell is aligned in a retro-reflected 6-beam MOT.}
\label{setup1}
\end{figure}

The MEMS vacuum cell is initially implemented in a 6-beam MOT set-up, shown in Fig.~\ref{setup1}. The cell consists of a $40\,\textrm{mm}\times20\,\textrm{mm}\times 4\,\textrm{mm}$ silicon frame with a $4\,$mm-wide rim at the bonding surfaces. The dimensions are chosen to provide a large surface bond area for increased probability of a hermetic seal, while providing a $12\,$mm inner dimension for optical access. The silicon is etched using a potassium hydroxide wet-etch through the complete $4\,$mm silicon thickness, via $2\,$mm-deep etches from both sides of the wafer. To avoid walls at angles defined by the crystal orientation, an over-etching technique is used to produce smooth inner wall surfaces, orthogonal to the upper and lower bonding surfaces \cite{Chutani2015}. 

The frame is anodically bonded to two $40\,\textrm{mm}\times 20\,\textrm{mm}$ glass wafers with 700$\,\mu$m thickness. The process of anodic bonding is usually carried out at a temperature near 300$^{\circ}$C to enable fast diffusion of the alkali ions within the glass. However, the use of glass doped with lithium instead of sodium or potassium allows anodic bonding at temperatures as low as 150$^{\circ}$C \cite{bondingtemp}. Low-temperature anodic bonding is required for MEMS vapor cells that are hindered by vapor diffusion into the glass cell windows at high temperatures. Reducing the bond temperature also alleviates the risk of damage to integrated components and circuitry \cite{lowtempbond}. In addition, we have found that the lithium-doped glass has a 6$\times$ larger initial bonding current and also bonds 8$\times$ faster compared to other comparable glass wafers (SD-2 Hoya \cite{Dellis2016}\footnote{Product name is for technical clarity; does not imply endorsement by NIST}) when bonded at 300$^{\circ}$C. A 6~mm-diameter hole is laser-cut in the upper ASG window through which the cell is connected to an ion-pump and resistive alkali-source. This method permits initial characterization of the cell hermeticity and provides the active-pumping currently required for laser cooling. Following the pump-down of the cell, a background pressure of 10$^{-9}$~mbar was achieved prior to alkali sourcing.

The MEMS cell is tilted with respect to the coil axis to obtain optical access for all 3 orthogonal counter-propagating beam pairs, with the overlap region placed inside the cell volume. An anti-Helmholtz coil pair is used to generate a $\approx 15\,$G/cm axial magnetic field within the overlap volume of the light field and MEMS cell.

The cooling and re-pumping light is derived from a single distributed-Bragg-reflector (DBR) laser, frequency modulated at $2.9\,$GHz to generate the required re-pumping light. The laser is frequency stabilized to the 780~nm cycling transition 5$S_{1/2}(F\!=\!3)\!\rightarrow$5$P_{3/2}(F'\!=\!4)$ with saturated-absorption spectroscopy. The cooling/repumping light is fiber-coupled and separately injected into the 6-beam and grating MOT systems, with maximum incident power of 20~mW and a red-detuning of $\delta$=-2$\Gamma$, where $\Gamma\approx2\pi\times 6\,$MHz is the transition's natural linewidth.

\section{Results}
The retro-reflected 6-beam MOT beams were aligned with $1/e^2$ diameter of 4~mm and approximately 1.5~mW in each arm. CCD imaging is aligned to avoid saturation of pixels from reflecting regions of the cell surfaces. When a steady-state Rb density is achieved from the resistively heated Rb$_2$MoO$_4$/AlZr alkali source, an atom number of 6.1(5)$\times$10$^5$ is observed from standard fluorescence imaging with a simultaneous rise-time extracted pressure of 1.1(5)$\times$10$^{-7}$~mbar. 

\begin{figure}[t]
\centering
\includegraphics[width=\columnwidth]{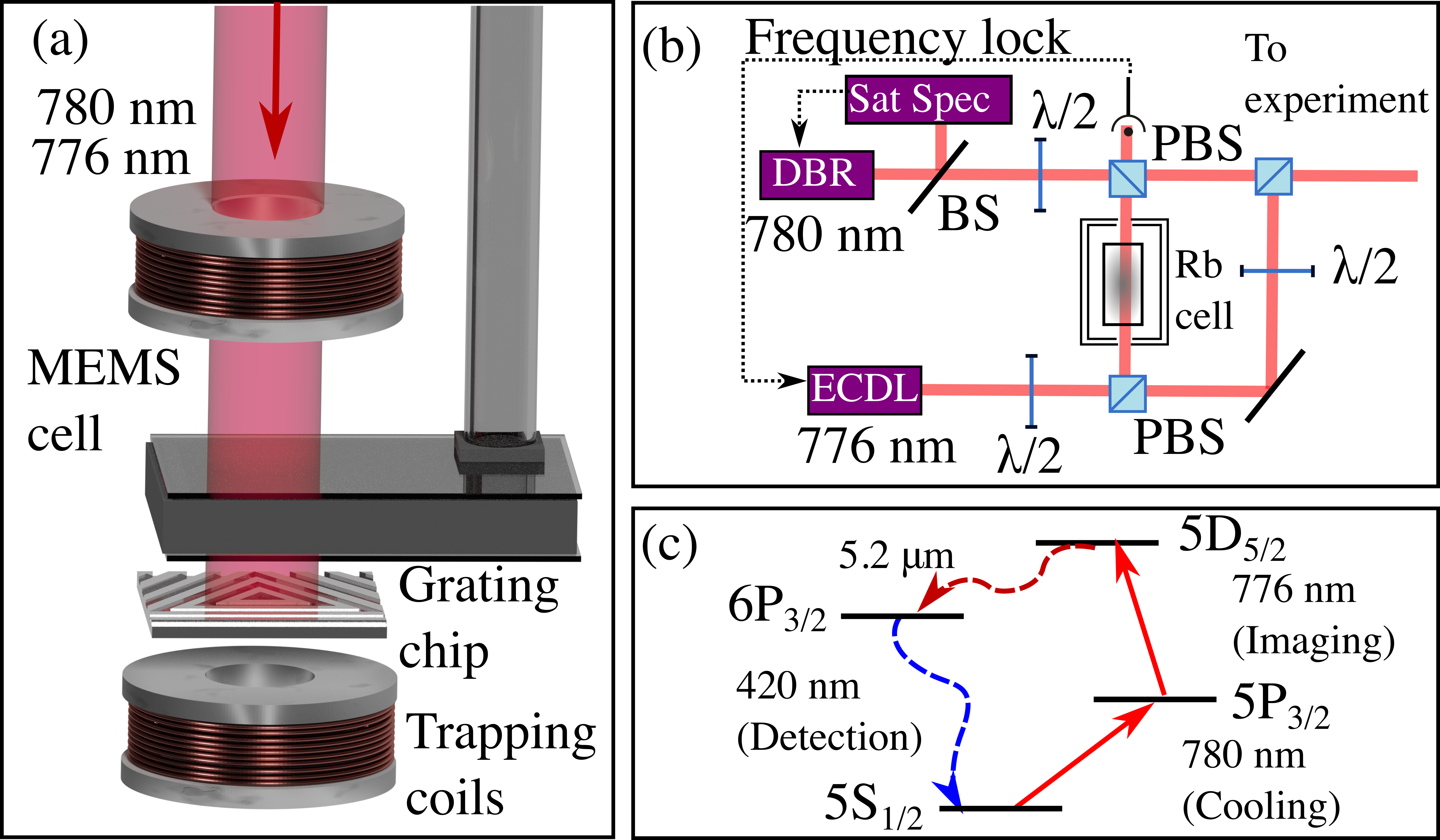}
\caption{(a): MEMS cell GMOT illustration. (b): Optics set-up for laser cooling and two-photon spectroscopy. (c): Coupled energy levels employed for blue fluorescence. BS: Beamsplitter. PBS: Polarizing beam splitter.}
\label{setup}
\end{figure}

To further simplify the cooling apparatus, the MEMS cell was implemented into a grating MOT package. Unlike the traditional 6-beam MOT, the GMOT requires no critical alignment of the MEMS cell for optical access, and can instead be placed above the grating chip in a planar stack, as shown in Fig.~\ref{setup} (a). Additionally, the cooling optics used for the 6-beam MOT are streamlined by implementation of the GMOT, providing atom cooling and trapping of $^{85}$Rb using a single incident beam, rather than three mutually-orthogonal counter-propagating beam pairs. The overlapped cooling (780~nm) beam and imaging (776~nm) beam are aligned with the center of the grating chip while the zeroth diffraction order is retro-reflected along the incident path.

A $1\,$cm$^2$ segmented diffraction grating is placed outside and immediately below the MEMS cell; to ensure that the laser overlap volume is well within the 4$\,$mm height of the cell. The smaller grating size used here compared to other GMOTs \cite{McGilligan2017,Nshii2013} was chosen to reduce the size constraints of the MEMS cell inner dimension. However, the smaller size leads to a reduced beam overlap volume $V$ and correspondingly lower steady-state trapped atom number $N\propto V^{1.2}$, implying an overlap volume ratio that is 11$\times$ smaller than the reported 2~cm$^2$ grating chips. The reduced overlap volume corresponds to a maximum achievable MOT atom number of about $10^6$ \cite{Nshii2013}.

To provide optical pumping for background-free imaging, an additional 776~nm laser was added to the optical set-up, illustrated in Fig.~\ref{setup} (b). The $776\,$nm light is derived from an external-cavity diode laser (ECDL \cite{Arnold1998}). The light is locked to the 5$D_{5/2}, F''=5$ state for maximum fluorescence by counter-propagating it with the 780~nm light through a 10~cm-long heated vapor cell, and detecting the ladder transition 5$S_{1/2}\rightarrow$5$P_{3/2}\rightarrow5D_{5/2}$, illustrated in Fig.~\ref{setup} (c), via the absorption of the 776~nm light. The remaining $776\,$nm light is combined with the cooling light on a polarizing-beam-splitter (PBS), and coupled into the same fiber as the cooling light to ensure good beam alignment. As a result of combining the beams on a PBS, both beams have orthogonal linear polarizations, later converted into orthogonal circular polarization using a $\lambda/4$ plate. The single incident beam was then expanded and apertured to a 1~cm diameter. 

Due to the $4\,$mm silicon walls, which are highly opaque to $420$~nm to $780$~nm light, a CCD was aligned at $\simeq$35$^{\circ}$ with respect to the grating chip plane to look through the upper cell window onto the grating chip surface. MOT images were acquired by imaging the blue fluorescence onto a CCD camera through a $420\,$nm band-pass filter to suppress the near-infra-red cooling light that is scattered from the surface of the grating chip. A background image, taken with the magnetic field gradient reversed, was subtracted from each MOT image to remove the small contribution of 420~nm light from the vapor and laboratory environment.

\begin{figure}[t]
\centering
\includegraphics[width=85mm]{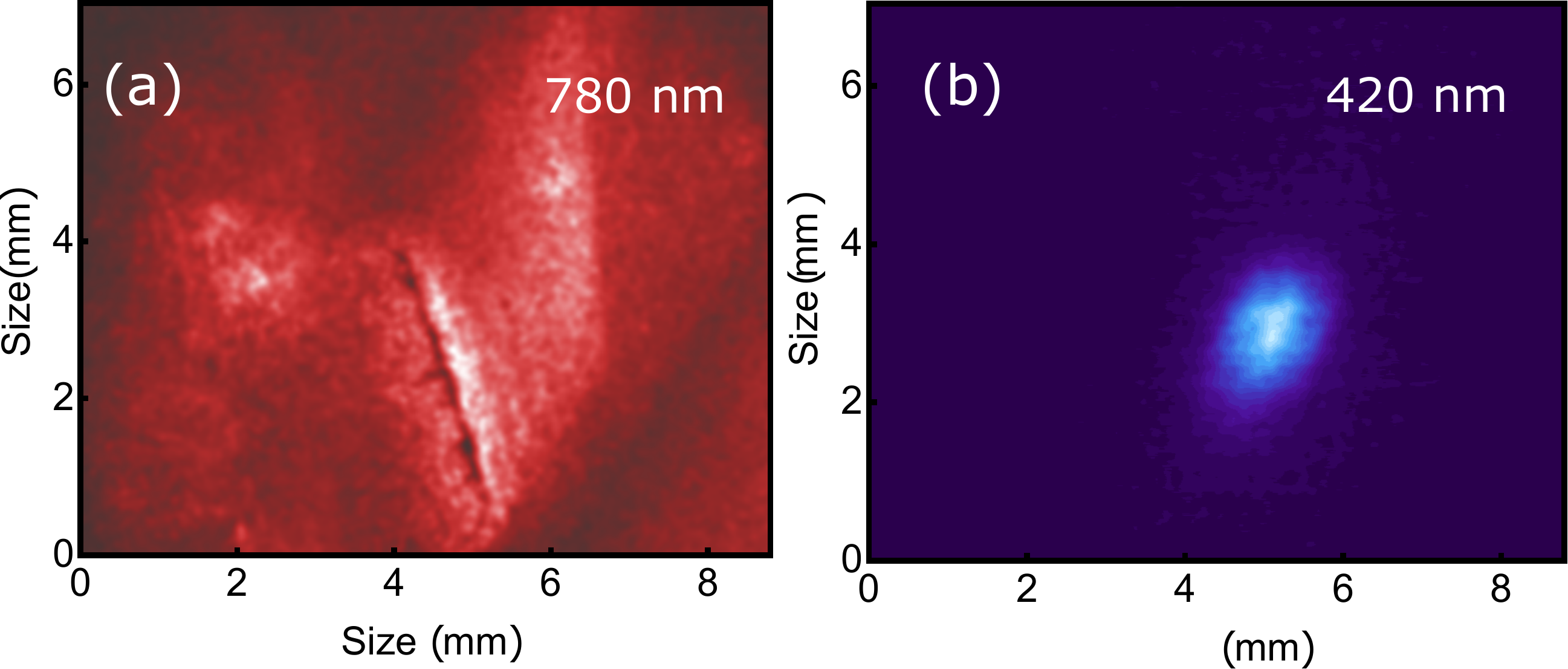}
\caption{Imaging of the GMOT (a): $780\,$nm fluorescence detection with $780\,$nm optics, $t_\textrm{exp}=5\,$ms. The surface scatter from the grating chip dominates the detection. (b): $420\,$nm fluorescence detection with a blue band-pass filter, $t_\textrm{exp}=1.5\,$s. A MOT is clearly visible with no background surface-scatter.}
\label{MOTimage}
\end{figure}

The initial attempts at imaging cold atoms in the chip-scale platform used $780\,$nm fluorescence imaging, shown in Fig.~\ref{MOTimage} (a). Under these conditions, the scattered light from the grating surface and the MEMS cell walls generates a large background, preventing cold-atom detection. However, when the $776\,$nm light is aligned and a $420\,$nm band-pass filter is added prior to the CCD, the only detected photons arise from atoms undergoing two-photon absorption in atomic Rb, removing all infra-red scattered light, as seen in Fig.~\ref{MOTimage} (b).

Background-free imaging of laser-cooled Rb via step-wise transitions 5$S_{1/2}(F\!=\!3)\!\rightarrow$5$P_{3/2}(F'\!=\!4)\!\rightarrow$5$D_{5/2}(F''\!=\!5)$ has been demonstrated previously \cite{Grove1995}. An incident wavelength of $\lambda_1=780\,$nm couples atoms through the cooling transition from the 5$S_{1/2}$ ground state to the 5$P_{3/2}(F'=4)$ excited state. A second laser at $\lambda_2=776\,$nm drives atoms to the 5$D_{5/2}(F''=5)$ excited state, where spontaneous emission permits decay back into the 5$P_{3/2}$ or 6$P_{3/2}$ state, with respective branching ratios of 0.65 and 0.35. A blue photon is generated when an atom decays from the 6$P_{3/2}$ levels to the 5$S_{1/2}$ ground state, with a branching ratio of 0.31. Atoms that are optically pumped to the 5$D_{5/2}$ state have an overall probability of 0.108 of generating a blue photon through the radiative cascade. 

Establishing the number of cold atoms from the blue fluorescence power is complicated by the effects of the 776~nm light on the laser cooling and trapping \cite{Grove1995}. For example, the 776~nm light causes optical pumping, which can reduce the fluorescence emitted per atom, and can also exert optical forces on the atoms, which can reduce the number of atoms in the trap. Due to these issues, direct atom number extraction has not been characterized from the MEMS grating system, however atom numbers of $5(1)\times 10^5$ were observed in a standard cuvette cell with this size of grating chip using incident laser powers of 20~mW. Extraction of the atom number directly from the two-photon scattering rate will be investigated further in the future.

\section{Discussion}
In this experiment, the vacuum inside the MEMS cell was maintained with active pumping by a 2~L/s ion pump, which dominates the volume of the overall system. The elimination of this larger element is clearly critical for further size reduction. It is possible that purely passive pumping with getters \cite{rushton,sebby2016} could be sufficient to maintain the required $\approx10^{-8}$ Torr vacuum if suitable materials are used to prevent the diffusion of noble gases into the cell \cite{Dellis2016}. It is also possible that more compact ion pumps could be developed, for example using silicon micro-machining processes \cite{wright, grzebyk}. If successful, these methods could be combined with the work described here to realize a highly compact and simple platform for cold-atom generation and instrumentation, which would allow for much wider deployment of cold-atom systems in real-world applications.

In summary, we demonstrated laser cooling from a traditional 6-beam MOT in a MEMS UHV cell, as well as the combination of MEMS UHV-cells and micro-fabricated diffractive optical elements to achieve laser-cooling in a planar-stacked device. The reduced optical access from the silicon cell walls restrains fluorescence imaging to non-zero angles with respect to the grating surface, such that scattered light greatly restricts the imaging quality. To circumvent this issue, a two-photon process in rubidium was used to achieve background-free imaging and resolve laser cooled atoms in the micro-fabricated package. Further investigation of the two-photon process is required. However, this method has permitted successful observation of cold atoms that overcomes surface-scatter in the chip-scale package. The micro-fabricated components used in conjunction with low incident laser powers suggests that this approach might be an ideal candidate for laser-cooling in low size-weight-and-power applications. 

\section{Acknowledgements}
The authors acknowledge J. W. Pollock and W. McGehee for careful reading of the manuscript before submission. The authors thank A.\ Hansen,
Y.-J.\ Chen and M.\ Shuker for fruitful discussions. R.B was supported by the NIST Guest Researcher program and D\'el\'egation G\'en\'erale de l'Armement (DGA). J.P.M. gratefully acknowledges support from the English Speaking Union and Lindemann Fellowship. G. D. M. was supported under the financial assistance award 70NANB18H006 from the U.S. Department of Commerce, National Institute of Standards and Technology. A.S.A, P.F.G, and E.R. gratefully acknowledge funding from EPSRC through grant EP/M013294/1 and EP/T001046/1 and Kelvin Nanotechnology for the fabrication of the non-standard grating chip.

\section{Additional Information}
The authors declare they have no competing interests. Approved for Public Release, Distribution Unlimited.

\bibliography{paper1}

\begin{thebibliography}{35}%
\makeatletter
\providecommand \@ifxundefined [1]{%
 \@ifx{#1\undefined}
}%
\providecommand \@ifnum [1]{%
 \ifnum #1\expandafter \@firstoftwo
 \else \expandafter \@secondoftwo
 \fi
}%
\providecommand \@ifx [1]{%
 \ifx #1\expandafter \@firstoftwo
 \else \expandafter \@secondoftwo
 \fi
}%
\providecommand \natexlab [1]{#1}%
\providecommand \enquote  [1]{``#1''}%
\providecommand \bibnamefont  [1]{#1}%
\providecommand \bibfnamefont [1]{#1}%
\providecommand \citenamefont [1]{#1}%
\providecommand \href@noop [0]{\@secondoftwo}%
\providecommand \href [0]{\begingroup \@sanitize@url \@href}%
\providecommand \@href[1]{\@@startlink{#1}\@@href}%
\providecommand \@@href[1]{\endgroup#1\@@endlink}%
\providecommand \@sanitize@url [0]{\catcode `\\12\catcode `\$12\catcode
  `\&12\catcode `\#12\catcode `\^12\catcode `\_12\catcode `\%12\relax}%
\providecommand \@@startlink[1]{}%
\providecommand \@@endlink[0]{}%
\providecommand \url  [0]{\begingroup\@sanitize@url \@url }%
\providecommand \@url [1]{\endgroup\@href {#1}{\urlprefix }}%
\providecommand \urlprefix  [0]{URL }%
\providecommand \Eprint [0]{\href }%
\providecommand \doibase [0]{http://dx.doi.org/}%
\providecommand \selectlanguage [0]{\@gobble}%
\providecommand \bibinfo  [0]{\@secondoftwo}%
\providecommand \bibfield  [0]{\@secondoftwo}%
\providecommand \translation [1]{[#1]}%
\providecommand \BibitemOpen [0]{}%
\providecommand \bibitemStop [0]{}%
\providecommand \bibitemNoStop [0]{.\EOS\space}%
\providecommand \EOS [0]{\spacefactor3000\relax}%
\providecommand \BibitemShut  [1]{\csname bibitem#1\endcsname}%
\let\auto@bib@innerbib\@empty
\bibitem [{\citenamefont {Hinkley}\ \emph {et~al.}(2013)\citenamefont
  {Hinkley}, \citenamefont {Sherman}, \citenamefont {Phillips}, \citenamefont
  {Schioppo}, \citenamefont {Lemke}, \citenamefont {Beloy}, \citenamefont
  {Pizzocaro}, \citenamefont {Oates},\ and\ \citenamefont
  {Ludlow}}]{Hinkley2013}%
  \BibitemOpen
  \bibfield  {author} {\bibinfo {author} {\bibfnamefont {N.}~\bibnamefont
  {Hinkley}}, \bibinfo {author} {\bibfnamefont {J.~A.}\ \bibnamefont
  {Sherman}}, \bibinfo {author} {\bibfnamefont {N.~B.}\ \bibnamefont
  {Phillips}}, \bibinfo {author} {\bibfnamefont {M.}~\bibnamefont {Schioppo}},
  \bibinfo {author} {\bibfnamefont {N.~D.}\ \bibnamefont {Lemke}}, \bibinfo
  {author} {\bibfnamefont {K.}~\bibnamefont {Beloy}}, \bibinfo {author}
  {\bibfnamefont {M.}~\bibnamefont {Pizzocaro}}, \bibinfo {author}
  {\bibfnamefont {C.~W.}\ \bibnamefont {Oates}}, \ and\ \bibinfo {author}
  {\bibfnamefont {A.~D.}\ \bibnamefont {Ludlow}},\ }\href {\doibase
  10.1126/science.1240420} {\bibfield  {journal} {\bibinfo  {journal}
  {Science}\ }\textbf {\bibinfo {volume} {341}},\ \bibinfo {pages} {1215}
  (\bibinfo {year} {2013})}\BibitemShut {NoStop}%
\bibitem [{\citenamefont {Kovachy}\ \emph {et~al.}(2015)\citenamefont
  {Kovachy}, \citenamefont {Asenbaum}, \citenamefont {Overstreet},
  \citenamefont {Donnelly}, \citenamefont {Dickerson}, \citenamefont
  {Sugarbaker}, \citenamefont {Hogan},\ and\ \citenamefont
  {Kasevich}}]{Kovachy2015}%
  \BibitemOpen
  \bibfield  {author} {\bibinfo {author} {\bibfnamefont {T.}~\bibnamefont
  {Kovachy}}, \bibinfo {author} {\bibfnamefont {P.}~\bibnamefont {Asenbaum}},
  \bibinfo {author} {\bibfnamefont {C.}~\bibnamefont {Overstreet}}, \bibinfo
  {author} {\bibfnamefont {C.~A.}\ \bibnamefont {Donnelly}}, \bibinfo {author}
  {\bibfnamefont {S.~M.}\ \bibnamefont {Dickerson}}, \bibinfo {author}
  {\bibfnamefont {A.}~\bibnamefont {Sugarbaker}}, \bibinfo {author}
  {\bibfnamefont {J.~M.}\ \bibnamefont {Hogan}}, \ and\ \bibinfo {author}
  {\bibfnamefont {M.~A.}\ \bibnamefont {Kasevich}},\ }\href {\doibase
  10.1038/nature16155} {\bibfield  {journal} {\bibinfo  {journal} {Nature}\
  }\textbf {\bibinfo {volume} {528}},\ \bibinfo {pages} {530} (\bibinfo {year}
  {2015})}\BibitemShut {NoStop}%
\bibitem [{\citenamefont {Laurent}\ \emph {et~al.}(1998)\citenamefont
  {Laurent}, \citenamefont {Lemonde}, \citenamefont {Simon}, \citenamefont
  {Santarelli}, \citenamefont {Clairon}, \citenamefont {Dimarcq}, \citenamefont
  {Petit}, \citenamefont {Audoin},\ and\ \citenamefont {Salomon}}]{Laurent}%
  \BibitemOpen
  \bibfield  {author} {\bibinfo {author} {\bibfnamefont {P.}~\bibnamefont
  {Laurent}}, \bibinfo {author} {\bibfnamefont {P.}~\bibnamefont {Lemonde}},
  \bibinfo {author} {\bibfnamefont {E.}~\bibnamefont {Simon}}, \bibinfo
  {author} {\bibfnamefont {G.}~\bibnamefont {Santarelli}}, \bibinfo {author}
  {\bibfnamefont {A.}~\bibnamefont {Clairon}}, \bibinfo {author} {\bibfnamefont
  {N.}~\bibnamefont {Dimarcq}}, \bibinfo {author} {\bibfnamefont
  {P.}~\bibnamefont {Petit}}, \bibinfo {author} {\bibfnamefont
  {C.}~\bibnamefont {Audoin}}, \ and\ \bibinfo {author} {\bibfnamefont
  {C.}~\bibnamefont {Salomon}},\ }\href {\doibase 10.1007/PL00021584}
  {\bibfield  {journal} {\bibinfo  {journal} {The European Physical Journal D -
  Atomic, Molecular, Optical and Plasma Physics}\ }\textbf {\bibinfo {volume}
  {3}},\ \bibinfo {pages} {201} (\bibinfo {year} {1998})}\BibitemShut {NoStop}%
\bibitem [{\citenamefont {Bongs}\ \emph {et~al.}(2019)\citenamefont {Bongs},
  \citenamefont {Holynski}, \citenamefont {Vovrosh}, \citenamefont {Bouyer},
  \citenamefont {Condon}, \citenamefont {Rasel}, \citenamefont {Schubert},
  \citenamefont {Schleich},\ and\ \citenamefont {Roura}}]{Bongs}%
  \BibitemOpen
  \bibfield  {author} {\bibinfo {author} {\bibfnamefont {K.}~\bibnamefont
  {Bongs}}, \bibinfo {author} {\bibfnamefont {M.}~\bibnamefont {Holynski}},
  \bibinfo {author} {\bibfnamefont {J.}~\bibnamefont {Vovrosh}}, \bibinfo
  {author} {\bibfnamefont {P.}~\bibnamefont {Bouyer}}, \bibinfo {author}
  {\bibfnamefont {G.}~\bibnamefont {Condon}}, \bibinfo {author} {\bibfnamefont
  {E.}~\bibnamefont {Rasel}}, \bibinfo {author} {\bibfnamefont
  {C.}~\bibnamefont {Schubert}}, \bibinfo {author} {\bibfnamefont {W.~P.}\
  \bibnamefont {Schleich}}, \ and\ \bibinfo {author} {\bibfnamefont
  {A.}~\bibnamefont {Roura}},\ }\href {\doibase 10.1038/s42254-019-0117-4}
  {\bibfield  {journal} {\bibinfo  {journal} {Nature Reviews Physics}\ }\textbf
  {\bibinfo {volume} {1}},\ \bibinfo {pages} {731} (\bibinfo {year}
  {2019})}\BibitemShut {NoStop}%
\bibitem [{\citenamefont {Knappe}\ \emph {et~al.}(2004)\citenamefont {Knappe},
  \citenamefont {Shah}, \citenamefont {Schwindt}, \citenamefont {Hollberg},
  \citenamefont {Kitching}, \citenamefont {Liew},\ and\ \citenamefont
  {Moreland}}]{Knappe2004}%
  \BibitemOpen
  \bibfield  {author} {\bibinfo {author} {\bibfnamefont {S.}~\bibnamefont
  {Knappe}}, \bibinfo {author} {\bibfnamefont {V.}~\bibnamefont {Shah}},
  \bibinfo {author} {\bibfnamefont {P.~D.~D.}\ \bibnamefont {Schwindt}},
  \bibinfo {author} {\bibfnamefont {L.}~\bibnamefont {Hollberg}}, \bibinfo
  {author} {\bibfnamefont {J.}~\bibnamefont {Kitching}}, \bibinfo {author}
  {\bibfnamefont {L.-A.}\ \bibnamefont {Liew}}, \ and\ \bibinfo {author}
  {\bibfnamefont {J.}~\bibnamefont {Moreland}},\ }\href {\doibase
  10.1063/1.1787942} {\bibfield  {journal} {\bibinfo  {journal} {Applied
  Physics Letters}\ }\textbf {\bibinfo {volume} {85}},\ \bibinfo {pages} {1460}
  (\bibinfo {year} {2004})}\BibitemShut {NoStop}%
\bibitem [{\citenamefont {Schwindt}\ \emph {et~al.}(2007)\citenamefont
  {Schwindt}, \citenamefont {Lindseth}, \citenamefont {Knappe}, \citenamefont
  {Shah}, \citenamefont {Kitching},\ and\ \citenamefont {Liew}}]{Schwindt2004}%
  \BibitemOpen
  \bibfield  {author} {\bibinfo {author} {\bibfnamefont {P.~D.~D.}\
  \bibnamefont {Schwindt}}, \bibinfo {author} {\bibfnamefont {B.}~\bibnamefont
  {Lindseth}}, \bibinfo {author} {\bibfnamefont {S.}~\bibnamefont {Knappe}},
  \bibinfo {author} {\bibfnamefont {V.}~\bibnamefont {Shah}}, \bibinfo {author}
  {\bibfnamefont {J.}~\bibnamefont {Kitching}}, \ and\ \bibinfo {author}
  {\bibfnamefont {L.-A.}\ \bibnamefont {Liew}},\ }\href {\doibase
  10.1063/1.2709532} {\bibfield  {journal} {\bibinfo  {journal} {Applied
  Physics Letters}\ }\textbf {\bibinfo {volume} {90}},\ \bibinfo {pages}
  {081102} (\bibinfo {year} {2007})}\BibitemShut {NoStop}%
\bibitem [{\citenamefont {Liew}\ \emph {et~al.}(2004)\citenamefont {Liew},
  \citenamefont {Knappe}, \citenamefont {Moreland}, \citenamefont {Robinson},
  \citenamefont {Hollberg},\ and\ \citenamefont {Kitching}}]{Liew2004}%
  \BibitemOpen
  \bibfield  {author} {\bibinfo {author} {\bibfnamefont {L.-A.}\ \bibnamefont
  {Liew}}, \bibinfo {author} {\bibfnamefont {S.}~\bibnamefont {Knappe}},
  \bibinfo {author} {\bibfnamefont {J.}~\bibnamefont {Moreland}}, \bibinfo
  {author} {\bibfnamefont {H.}~\bibnamefont {Robinson}}, \bibinfo {author}
  {\bibfnamefont {L.}~\bibnamefont {Hollberg}}, \ and\ \bibinfo {author}
  {\bibfnamefont {J.}~\bibnamefont {Kitching}},\ }\href {\doibase
  10.1063/1.1691490} {\bibfield  {journal} {\bibinfo  {journal} {Applied
  Physics Letters}\ }\textbf {\bibinfo {volume} {84}},\ \bibinfo {pages} {2694}
  (\bibinfo {year} {2004})}\BibitemShut {NoStop}%
\bibitem [{\citenamefont {Kitching}(2018)}]{Kitching2018}%
  \BibitemOpen
  \bibfield  {author} {\bibinfo {author} {\bibfnamefont {J.}~\bibnamefont
  {Kitching}},\ }\href {\doibase 10.1063/1.5026238} {\bibfield  {journal}
  {\bibinfo  {journal} {Applied Physics Reviews}\ }\textbf {\bibinfo {volume}
  {5}},\ \bibinfo {pages} {031302} (\bibinfo {year} {2018})}\BibitemShut
  {NoStop}%
\bibitem [{\citenamefont {Trupke}\ \emph {et~al.}(2006)\citenamefont {Trupke},
  \citenamefont {Ramirez-Martinez}, \citenamefont {Curtis}, \citenamefont
  {Ashmore}, \citenamefont {Eriksson}, \citenamefont {Hinds}, \citenamefont
  {Moktadir}, \citenamefont {Gollasch}, \citenamefont {Kraft}, \citenamefont
  {Vijaya~Prakash},\ and\ \citenamefont {Baumberg}}]{Trupte2006}%
  \BibitemOpen
  \bibfield  {author} {\bibinfo {author} {\bibfnamefont {M.}~\bibnamefont
  {Trupke}}, \bibinfo {author} {\bibfnamefont {F.}~\bibnamefont
  {Ramirez-Martinez}}, \bibinfo {author} {\bibfnamefont {E.~A.}\ \bibnamefont
  {Curtis}}, \bibinfo {author} {\bibfnamefont {J.~P.}\ \bibnamefont {Ashmore}},
  \bibinfo {author} {\bibfnamefont {S.}~\bibnamefont {Eriksson}}, \bibinfo
  {author} {\bibfnamefont {E.~A.}\ \bibnamefont {Hinds}}, \bibinfo {author}
  {\bibfnamefont {Z.}~\bibnamefont {Moktadir}}, \bibinfo {author}
  {\bibfnamefont {C.}~\bibnamefont {Gollasch}}, \bibinfo {author}
  {\bibfnamefont {M.}~\bibnamefont {Kraft}}, \bibinfo {author} {\bibfnamefont
  {G.}~\bibnamefont {Vijaya~Prakash}}, \ and\ \bibinfo {author} {\bibfnamefont
  {J.~J.}\ \bibnamefont {Baumberg}},\ }\href {\doibase 10.1063/1.2172412}
  {\bibfield  {journal} {\bibinfo  {journal} {Applied Physics Letters}\
  }\textbf {\bibinfo {volume} {88}},\ \bibinfo {pages} {071116} (\bibinfo
  {year} {2006})}\BibitemShut {NoStop}%
\bibitem [{\citenamefont {Vangeleyn}\ \emph {et~al.}(2009)\citenamefont
  {Vangeleyn}, \citenamefont {Griffin}, \citenamefont {Riis},\ and\
  \citenamefont {Arnold}}]{arnold1}%
  \BibitemOpen
  \bibfield  {author} {\bibinfo {author} {\bibfnamefont {M.}~\bibnamefont
  {Vangeleyn}}, \bibinfo {author} {\bibfnamefont {P.~F.}\ \bibnamefont
  {Griffin}}, \bibinfo {author} {\bibfnamefont {E.}~\bibnamefont {Riis}}, \
  and\ \bibinfo {author} {\bibfnamefont {A.~S.}\ \bibnamefont {Arnold}},\
  }\href {\doibase 10.1364/OE.17.013601} {\bibfield  {journal} {\bibinfo
  {journal} {Optics Express}\ }\textbf {\bibinfo {volume} {17}},\ \bibinfo
  {pages} {13601} (\bibinfo {year} {2009})}\BibitemShut {NoStop}%
\bibitem [{\citenamefont {Lee}\ \emph {et~al.}(1996)\citenamefont {Lee},
  \citenamefont {Kim}, \citenamefont {Noh},\ and\ \citenamefont {Jhe}}]{Lee96}%
  \BibitemOpen
  \bibfield  {author} {\bibinfo {author} {\bibfnamefont {K.~I.}\ \bibnamefont
  {Lee}}, \bibinfo {author} {\bibfnamefont {J.~A.}\ \bibnamefont {Kim}},
  \bibinfo {author} {\bibfnamefont {H.~R.}\ \bibnamefont {Noh}}, \ and\
  \bibinfo {author} {\bibfnamefont {W.}~\bibnamefont {Jhe}},\ }\href {\doibase
  10.1364/OL.21.001177} {\bibfield  {journal} {\bibinfo  {journal} {Optics
  Express}\ }\textbf {\bibinfo {volume} {21}},\ \bibinfo {pages} {1177}
  (\bibinfo {year} {1996})}\BibitemShut {NoStop}%
\bibitem [{\citenamefont {Shah}\ \emph {et~al.}(2011)\citenamefont {Shah},
  \citenamefont {Mescher}, \citenamefont {Martins}, \citenamefont {Leblanc},
  \citenamefont {Byrne}, \citenamefont {Timmons}, \citenamefont {Stoner},
  \citenamefont {Rogamentich},\ and\ \citenamefont {Lutwak}}]{Shah2011}%
  \BibitemOpen
  \bibfield  {author} {\bibinfo {author} {\bibfnamefont {V.}~\bibnamefont
  {Shah}}, \bibinfo {author} {\bibfnamefont {M.}~\bibnamefont {Mescher}},
  \bibinfo {author} {\bibfnamefont {A.}~\bibnamefont {Martins}}, \bibinfo
  {author} {\bibfnamefont {J.}~\bibnamefont {Leblanc}}, \bibinfo {author}
  {\bibfnamefont {N.}~\bibnamefont {Byrne}}, \bibinfo {author} {\bibfnamefont
  {B.}~\bibnamefont {Timmons}}, \bibinfo {author} {\bibfnamefont
  {R.}~\bibnamefont {Stoner}}, \bibinfo {author} {\bibfnamefont
  {F.}~\bibnamefont {Rogamentich}}, \ and\ \bibinfo {author} {\bibfnamefont
  {R.}~\bibnamefont {Lutwak}},\ }\href@noop {} {\bibfield  {journal} {\bibinfo
  {journal} {Proceedings of the 43rd Annual Precise Time and Time Interval
  Systems and Applications Meeting, Long Beach, California}\ ,\ \bibinfo
  {pages} {221}} (\bibinfo {year} {2011})}\BibitemShut {NoStop}%
\bibitem [{\citenamefont {Sebby-Strabley}\ \emph {et~al.}(2016)\citenamefont
  {Sebby-Strabley}, \citenamefont {Fertig}, \citenamefont {Compton},
  \citenamefont {Salit}, \citenamefont {Nelson}, \citenamefont {Stark},
  \citenamefont {Langness},\ and\ \citenamefont {Livingston}}]{sebby2016}%
  \BibitemOpen
  \bibfield  {author} {\bibinfo {author} {\bibfnamefont {J.}~\bibnamefont
  {Sebby-Strabley}}, \bibinfo {author} {\bibfnamefont {C.}~\bibnamefont
  {Fertig}}, \bibinfo {author} {\bibfnamefont {R.}~\bibnamefont {Compton}},
  \bibinfo {author} {\bibfnamefont {K.}~\bibnamefont {Salit}}, \bibinfo
  {author} {\bibfnamefont {K.}~\bibnamefont {Nelson}}, \bibinfo {author}
  {\bibfnamefont {T.}~\bibnamefont {Stark}}, \bibinfo {author} {\bibfnamefont
  {C.}~\bibnamefont {Langness}}, \ and\ \bibinfo {author} {\bibfnamefont
  {R.}~\bibnamefont {Livingston}}\ }(\bibinfo {year} {2016})\ pp.\ \bibinfo
  {pages} {1--6}\BibitemShut {NoStop}%
\bibitem [{\citenamefont {Salim}\ \emph {et~al.}(2011)\citenamefont {Salim},
  \citenamefont {DeNatale}, \citenamefont {Farkas}, \citenamefont {Hudek},
  \citenamefont {McBride}, \citenamefont {Michalchuk}, \citenamefont
  {Mihailovich},\ and\ \citenamefont {Anderson}}]{selim}%
  \BibitemOpen
  \bibfield  {author} {\bibinfo {author} {\bibfnamefont {E.~A.}\ \bibnamefont
  {Salim}}, \bibinfo {author} {\bibfnamefont {J.}~\bibnamefont {DeNatale}},
  \bibinfo {author} {\bibfnamefont {D.~M.}\ \bibnamefont {Farkas}}, \bibinfo
  {author} {\bibfnamefont {K.~M.}\ \bibnamefont {Hudek}}, \bibinfo {author}
  {\bibfnamefont {S.~E.}\ \bibnamefont {McBride}}, \bibinfo {author}
  {\bibfnamefont {J.}~\bibnamefont {Michalchuk}}, \bibinfo {author}
  {\bibfnamefont {R.}~\bibnamefont {Mihailovich}}, \ and\ \bibinfo {author}
  {\bibfnamefont {D.~Z.}\ \bibnamefont {Anderson}},\ }\href {\doibase
  10.1007/s11128-011-0300-8} {\bibfield  {journal} {\bibinfo  {journal}
  {Quantum Information Processing}\ }\textbf {\bibinfo {volume} {10}},\
  \bibinfo {pages} {975} (\bibinfo {year} {2011})}\BibitemShut {NoStop}%
\bibitem [{\citenamefont {Reichel}\ \emph {et~al.}(1999)\citenamefont
  {Reichel}, \citenamefont {H\"ansel},\ and\ \citenamefont
  {H\"ansch}}]{Reichel1999}%
  \BibitemOpen
  \bibfield  {author} {\bibinfo {author} {\bibfnamefont {J.}~\bibnamefont
  {Reichel}}, \bibinfo {author} {\bibfnamefont {W.}~\bibnamefont {H\"ansel}}, \
  and\ \bibinfo {author} {\bibfnamefont {T.~W.}\ \bibnamefont {H\"ansch}},\
  }\href {\doibase 10.1103/PhysRevLett.83.3398} {\bibfield  {journal} {\bibinfo
   {journal} {Physical Review Letters}\ }\textbf {\bibinfo {volume} {83}},\
  \bibinfo {pages} {3398} (\bibinfo {year} {1999})}\BibitemShut {NoStop}%
\bibitem [{\citenamefont {Folman}\ \emph {et~al.}(2000)\citenamefont {Folman},
  \citenamefont {Kr\"uger}, \citenamefont {Cassettari}, \citenamefont {Hessmo},
  \citenamefont {Maier},\ and\ \citenamefont {Schmiedmayer}}]{Folman}%
  \BibitemOpen
  \bibfield  {author} {\bibinfo {author} {\bibfnamefont {R.}~\bibnamefont
  {Folman}}, \bibinfo {author} {\bibfnamefont {P.}~\bibnamefont {Kr\"uger}},
  \bibinfo {author} {\bibfnamefont {D.}~\bibnamefont {Cassettari}}, \bibinfo
  {author} {\bibfnamefont {B.}~\bibnamefont {Hessmo}}, \bibinfo {author}
  {\bibfnamefont {T.}~\bibnamefont {Maier}}, \ and\ \bibinfo {author}
  {\bibfnamefont {J.}~\bibnamefont {Schmiedmayer}},\ }\href {\doibase
  10.1103/PhysRevLett.84.4749} {\bibfield  {journal} {\bibinfo  {journal}
  {Physical Review Letters}\ }\textbf {\bibinfo {volume} {84}},\ \bibinfo
  {pages} {4749} (\bibinfo {year} {2000})}\BibitemShut {NoStop}%
\bibitem [{\citenamefont {Ram{\'\i}rez-Mart{\'\i}nez}\ \emph
  {et~al.}(2011)\citenamefont {Ram{\'\i}rez-Mart{\'\i}nez}, \citenamefont
  {Lacro{\^u}te}, \citenamefont {Rosenbusch}, \citenamefont {Reinhard},
  \citenamefont {Deutsch}, \citenamefont {Schneider},\ and\ \citenamefont
  {Reichel}}]{ramirez}%
  \BibitemOpen
  \bibfield  {author} {\bibinfo {author} {\bibfnamefont {F.}~\bibnamefont
  {Ram{\'\i}rez-Mart{\'\i}nez}}, \bibinfo {author} {\bibfnamefont
  {C.}~\bibnamefont {Lacro{\^u}te}}, \bibinfo {author} {\bibfnamefont
  {P.}~\bibnamefont {Rosenbusch}}, \bibinfo {author} {\bibfnamefont
  {F.}~\bibnamefont {Reinhard}}, \bibinfo {author} {\bibfnamefont
  {C.}~\bibnamefont {Deutsch}}, \bibinfo {author} {\bibfnamefont
  {T.}~\bibnamefont {Schneider}}, \ and\ \bibinfo {author} {\bibfnamefont
  {J.}~\bibnamefont {Reichel}},\ }\bibfield  {booktitle} {\emph {\bibinfo
  {booktitle} {Scientific applications of Galileo and other Global Navigation
  Satellite Systems - I}},\ }\href {\doibase
  https://doi.org/10.1016/j.asr.2010.04.014} {\bibfield  {journal} {\bibinfo
  {journal} {Advances in Space Research}\ }\textbf {\bibinfo {volume} {47}},\
  \bibinfo {pages} {247} (\bibinfo {year} {2011})}\BibitemShut {NoStop}%
\bibitem [{\citenamefont {McGilligan}\ \emph {et~al.}(2017)\citenamefont
  {McGilligan}, \citenamefont {Griffin}, \citenamefont {Elvin}, \citenamefont
  {Ingleby}, \citenamefont {Riis},\ and\ \citenamefont
  {Arnold}}]{McGilligan2017}%
  \BibitemOpen
  \bibfield  {author} {\bibinfo {author} {\bibfnamefont {J.~P.}\ \bibnamefont
  {McGilligan}}, \bibinfo {author} {\bibfnamefont {P.~F.}\ \bibnamefont
  {Griffin}}, \bibinfo {author} {\bibfnamefont {R.}~\bibnamefont {Elvin}},
  \bibinfo {author} {\bibfnamefont {S.~J.}\ \bibnamefont {Ingleby}}, \bibinfo
  {author} {\bibfnamefont {E.}~\bibnamefont {Riis}}, \ and\ \bibinfo {author}
  {\bibfnamefont {A.~S.}\ \bibnamefont {Arnold}},\ }\href
  {https://doi.org/10.1038/s41598-017-00254-0} {\bibfield  {journal} {\bibinfo
  {journal} {Scientific Reports}\ }\textbf {\bibinfo {volume} {7}} (\bibinfo
  {year} {2017})}\BibitemShut {NoStop}%
\bibitem [{\citenamefont {Kang}\ \emph {et~al.}(2019)\citenamefont {Kang},
  \citenamefont {Moore}, \citenamefont {McGilligan}, \citenamefont {Mott},
  \citenamefont {Mis}, \citenamefont {Roper}, \citenamefont {Donley},\ and\
  \citenamefont {Kitching}}]{Kang2019}%
  \BibitemOpen
  \bibfield  {author} {\bibinfo {author} {\bibfnamefont {S.}~\bibnamefont
  {Kang}}, \bibinfo {author} {\bibfnamefont {K.~R.}\ \bibnamefont {Moore}},
  \bibinfo {author} {\bibfnamefont {J.~P.}\ \bibnamefont {McGilligan}},
  \bibinfo {author} {\bibfnamefont {R.}~\bibnamefont {Mott}}, \bibinfo {author}
  {\bibfnamefont {A.}~\bibnamefont {Mis}}, \bibinfo {author} {\bibfnamefont
  {C.}~\bibnamefont {Roper}}, \bibinfo {author} {\bibfnamefont {E.~A.}\
  \bibnamefont {Donley}}, \ and\ \bibinfo {author} {\bibfnamefont
  {J.}~\bibnamefont {Kitching}},\ }\href {\doibase 10.1364/ol.44.003002}
  {\bibfield  {journal} {\bibinfo  {journal} {Optics Letters}\ }\textbf
  {\bibinfo {volume} {44}},\ \bibinfo {pages} {3002} (\bibinfo {year}
  {2019})}\BibitemShut {NoStop}%
\bibitem [{\citenamefont {McGilligan}\ \emph {et~al.}(2020)\citenamefont
  {McGilligan}, \citenamefont {Moore}, \citenamefont {Kang}, \citenamefont
  {Mott}, \citenamefont {Mis}, \citenamefont {Roper}, \citenamefont {Donley},\
  and\ \citenamefont {Kitching}}]{mcgilliganPRApplied}%
  \BibitemOpen
  \bibfield  {author} {\bibinfo {author} {\bibfnamefont {J.~P.}\ \bibnamefont
  {McGilligan}}, \bibinfo {author} {\bibfnamefont {K.~R.}\ \bibnamefont
  {Moore}}, \bibinfo {author} {\bibfnamefont {S.}~\bibnamefont {Kang}},
  \bibinfo {author} {\bibfnamefont {R.}~\bibnamefont {Mott}}, \bibinfo {author}
  {\bibfnamefont {A.}~\bibnamefont {Mis}}, \bibinfo {author} {\bibfnamefont
  {C.}~\bibnamefont {Roper}}, \bibinfo {author} {\bibfnamefont {E.~A.}\
  \bibnamefont {Donley}}, \ and\ \bibinfo {author} {\bibfnamefont
  {J.}~\bibnamefont {Kitching}},\ }\href
  {https://link.aps.org/doi/10.1103/PhysRevApplied.13.044038} {\bibfield
  {journal} {\bibinfo  {journal} {Physical Review Applied}\ }\textbf {\bibinfo
  {volume} {13}},\ \bibinfo {pages} {044038} (\bibinfo {year}
  {2020})}\BibitemShut {NoStop}%
\bibitem [{\citenamefont {Saint}\ \emph {et~al.}(2018)\citenamefont {Saint},
  \citenamefont {Evans}, \citenamefont {Zhou}, \citenamefont {Barrett},
  \citenamefont {Fromhold}, \citenamefont {Saleh}, \citenamefont {Maskery},
  \citenamefont {Tuck}, \citenamefont {Wildman}, \citenamefont
  {Oru{\v{c}}evi{\'{c}}},\ and\ \citenamefont {Kr\"{u}ger}}]{Saint2018}%
  \BibitemOpen
  \bibfield  {author} {\bibinfo {author} {\bibfnamefont {R.}~\bibnamefont
  {Saint}}, \bibinfo {author} {\bibfnamefont {W.}~\bibnamefont {Evans}},
  \bibinfo {author} {\bibfnamefont {Y.}~\bibnamefont {Zhou}}, \bibinfo {author}
  {\bibfnamefont {T.}~\bibnamefont {Barrett}}, \bibinfo {author} {\bibfnamefont
  {T.~M.}\ \bibnamefont {Fromhold}}, \bibinfo {author} {\bibfnamefont
  {E.}~\bibnamefont {Saleh}}, \bibinfo {author} {\bibfnamefont
  {I.}~\bibnamefont {Maskery}}, \bibinfo {author} {\bibfnamefont
  {C.}~\bibnamefont {Tuck}}, \bibinfo {author} {\bibfnamefont {R.}~\bibnamefont
  {Wildman}}, \bibinfo {author} {\bibfnamefont {F.}~\bibnamefont
  {Oru{\v{c}}evi{\'{c}}}}, \ and\ \bibinfo {author} {\bibfnamefont
  {P.}~\bibnamefont {Kr\"{u}ger}},\ }\href
  {https://doi.org/10.1038/s41598-018-26455-9} {\bibfield  {journal} {\bibinfo
  {journal} {Scientific Reports}\ }\textbf {\bibinfo {volume} {8}} (\bibinfo
  {year} {2018})}\BibitemShut {NoStop}%
\bibitem [{\citenamefont {Chuang}\ and\ \citenamefont
  {Huang}(2014)}]{siliconbulk}%
  \BibitemOpen
  \bibfield  {author} {\bibinfo {author} {\bibfnamefont {H.-C.}\ \bibnamefont
  {Chuang}}\ and\ \bibinfo {author} {\bibfnamefont {C.-S.}\ \bibnamefont
  {Huang}},\ }\href {\doibase 10.1063/1.4879115} {\bibfield  {journal}
  {\bibinfo  {journal} {Review of Scientific Instruments}\ }\textbf {\bibinfo
  {volume} {85}},\ \bibinfo {pages} {053107} (\bibinfo {year}
  {2014})}\BibitemShut {NoStop}%
\bibitem [{\citenamefont {Dellis}\ \emph {et~al.}(2016)\citenamefont {Dellis},
  \citenamefont {Shah}, \citenamefont {Donley}, \citenamefont {Knappe},\ and\
  \citenamefont {Kitching}}]{Dellis2016}%
  \BibitemOpen
  \bibfield  {author} {\bibinfo {author} {\bibfnamefont {A.~T.}\ \bibnamefont
  {Dellis}}, \bibinfo {author} {\bibfnamefont {V.}~\bibnamefont {Shah}},
  \bibinfo {author} {\bibfnamefont {E.~A.}\ \bibnamefont {Donley}}, \bibinfo
  {author} {\bibfnamefont {S.}~\bibnamefont {Knappe}}, \ and\ \bibinfo {author}
  {\bibfnamefont {J.}~\bibnamefont {Kitching}},\ }\href {\doibase
  10.1364/ol.41.002775} {\bibfield  {journal} {\bibinfo  {journal} {Optics
  Letters}\ }\textbf {\bibinfo {volume} {41}},\ \bibinfo {pages} {2775}
  (\bibinfo {year} {2016})}\BibitemShut {NoStop}%
\bibitem [{\citenamefont {Ohadi}\ \emph {et~al.}(2009)\citenamefont {Ohadi},
  \citenamefont {Himsworth}, \citenamefont {Xuereb},\ and\ \citenamefont
  {Freegarde}}]{Ohadi2009}%
  \BibitemOpen
  \bibfield  {author} {\bibinfo {author} {\bibfnamefont {H.}~\bibnamefont
  {Ohadi}}, \bibinfo {author} {\bibfnamefont {M.}~\bibnamefont {Himsworth}},
  \bibinfo {author} {\bibfnamefont {A.}~\bibnamefont {Xuereb}}, \ and\ \bibinfo
  {author} {\bibfnamefont {T.}~\bibnamefont {Freegarde}},\ }\href {\doibase
  10.1364/oe.17.023003} {\bibfield  {journal} {\bibinfo  {journal} {Optics
  Express}\ }\textbf {\bibinfo {volume} {17}},\ \bibinfo {pages} {23003}
  (\bibinfo {year} {2009})}\BibitemShut {NoStop}%
\bibitem [{\citenamefont {Sheludko}\ \emph {et~al.}(2008)\citenamefont
  {Sheludko}, \citenamefont {Bell}, \citenamefont {Anderson}, \citenamefont
  {Hofmann}, \citenamefont {Vredenbregt},\ and\ \citenamefont
  {Scholten}}]{Sheludko2008}%
  \BibitemOpen
  \bibfield  {author} {\bibinfo {author} {\bibfnamefont {D.~V.}\ \bibnamefont
  {Sheludko}}, \bibinfo {author} {\bibfnamefont {S.~C.}\ \bibnamefont {Bell}},
  \bibinfo {author} {\bibfnamefont {R.}~\bibnamefont {Anderson}}, \bibinfo
  {author} {\bibfnamefont {C.~S.}\ \bibnamefont {Hofmann}}, \bibinfo {author}
  {\bibfnamefont {E.~J.~D.}\ \bibnamefont {Vredenbregt}}, \ and\ \bibinfo
  {author} {\bibfnamefont {R.~E.}\ \bibnamefont {Scholten}},\ }\href
  {https://doi.org/10.1103/physreva.77.033401} {\bibfield  {journal} {\bibinfo
  {journal} {Physical Review A}\ }\textbf {\bibinfo {volume} {77}} (\bibinfo
  {year} {2008})}\BibitemShut {NoStop}%
\bibitem [{\citenamefont {Chutani}\ \emph {et~al.}(2015)\citenamefont
  {Chutani}, \citenamefont {Maurice}, \citenamefont {Passilly}, \citenamefont
  {Gorecki}, \citenamefont {Boudot}, \citenamefont {Abdel-Hafiz}, \citenamefont
  {Abb{\'{e}}}, \citenamefont {Galliou}, \citenamefont {Rauch},\ and\
  \citenamefont {de~Clercq}}]{Chutani2015}%
  \BibitemOpen
  \bibfield  {author} {\bibinfo {author} {\bibfnamefont {R.}~\bibnamefont
  {Chutani}}, \bibinfo {author} {\bibfnamefont {V.}~\bibnamefont {Maurice}},
  \bibinfo {author} {\bibfnamefont {N.}~\bibnamefont {Passilly}}, \bibinfo
  {author} {\bibfnamefont {C.}~\bibnamefont {Gorecki}}, \bibinfo {author}
  {\bibfnamefont {R.}~\bibnamefont {Boudot}}, \bibinfo {author} {\bibfnamefont
  {M.}~\bibnamefont {Abdel-Hafiz}}, \bibinfo {author} {\bibfnamefont
  {P.}~\bibnamefont {Abb{\'{e}}}}, \bibinfo {author} {\bibfnamefont
  {S.}~\bibnamefont {Galliou}}, \bibinfo {author} {\bibfnamefont {J.-Y.}\
  \bibnamefont {Rauch}}, \ and\ \bibinfo {author} {\bibfnamefont
  {E.}~\bibnamefont {de~Clercq}},\ }\href {https://doi.org/10.1038/srep14001}
  {\bibfield  {journal} {\bibinfo  {journal} {Scientific Reports}\ }\textbf
  {\bibinfo {volume} {5}} (\bibinfo {year} {2015})}\BibitemShut {NoStop}%
\bibitem [{\citenamefont {Shoji}\ \emph {et~al.}(1998)\citenamefont {Shoji},
  \citenamefont {Kikuchi},\ and\ \citenamefont {Torigoe}}]{bondingtemp}%
  \BibitemOpen
  \bibfield  {author} {\bibinfo {author} {\bibfnamefont {S.}~\bibnamefont
  {Shoji}}, \bibinfo {author} {\bibfnamefont {H.}~\bibnamefont {Kikuchi}}, \
  and\ \bibinfo {author} {\bibfnamefont {H.}~\bibnamefont {Torigoe}},\ }\href
  {\doibase https://doi.org/10.1016/S0924-4247(97)01659-2} {\bibfield
  {journal} {\bibinfo  {journal} {Sensors and Actuators A: Physical}\ }\textbf
  {\bibinfo {volume} {64}},\ \bibinfo {pages} {95 } (\bibinfo {year}
  {1998})}\BibitemShut {NoStop}%
\bibitem [{\citenamefont {Vesborg}\ \emph {et~al.}(2010)\citenamefont
  {Vesborg}, \citenamefont {Olsen}, \citenamefont {Henriksen}, \citenamefont
  {Chorkendorff},\ and\ \citenamefont {Hansen}}]{lowtempbond}%
  \BibitemOpen
  \bibfield  {author} {\bibinfo {author} {\bibfnamefont {P.~C.~K.}\
  \bibnamefont {Vesborg}}, \bibinfo {author} {\bibfnamefont {J.~L.}\
  \bibnamefont {Olsen}}, \bibinfo {author} {\bibfnamefont {T.~R.}\ \bibnamefont
  {Henriksen}}, \bibinfo {author} {\bibfnamefont {I.}~\bibnamefont
  {Chorkendorff}}, \ and\ \bibinfo {author} {\bibfnamefont {O.}~\bibnamefont
  {Hansen}},\ }\href {\doibase 10.1063/1.3277117} {\bibfield  {journal}
  {\bibinfo  {journal} {Review of Scientific Instruments}\ }\textbf {\bibinfo
  {volume} {81}},\ \bibinfo {pages} {016111} (\bibinfo {year}
  {2010})}\BibitemShut {NoStop}%
\bibitem [{Note1()}]{Note1}%
  \BibitemOpen
  \bibinfo {note} {Product name is for technical clarity; does not imply
  endorsement by NIST}\BibitemShut {NoStop}%
\bibitem [{\citenamefont {Nshii}\ \emph {et~al.}(2013)\citenamefont {Nshii},
  \citenamefont {Vangeleyn}, \citenamefont {Cotter}, \citenamefont {Griffin},
  \citenamefont {Hinds}, \citenamefont {Ironside}, \citenamefont {See},
  \citenamefont {Sinclair}, \citenamefont {Riis},\ and\ \citenamefont
  {Arnold}}]{Nshii2013}%
  \BibitemOpen
  \bibfield  {author} {\bibinfo {author} {\bibfnamefont {C.~C.}\ \bibnamefont
  {Nshii}}, \bibinfo {author} {\bibfnamefont {M.}~\bibnamefont {Vangeleyn}},
  \bibinfo {author} {\bibfnamefont {J.~P.}\ \bibnamefont {Cotter}}, \bibinfo
  {author} {\bibfnamefont {P.~F.}\ \bibnamefont {Griffin}}, \bibinfo {author}
  {\bibfnamefont {E.~A.}\ \bibnamefont {Hinds}}, \bibinfo {author}
  {\bibfnamefont {C.~N.}\ \bibnamefont {Ironside}}, \bibinfo {author}
  {\bibfnamefont {P.}~\bibnamefont {See}}, \bibinfo {author} {\bibfnamefont
  {A.~G.}\ \bibnamefont {Sinclair}}, \bibinfo {author} {\bibfnamefont
  {E.}~\bibnamefont {Riis}}, \ and\ \bibinfo {author} {\bibfnamefont {A.~S.}\
  \bibnamefont {Arnold}},\ }\href {\doibase 10.1038/nnano.2013.47} {\bibfield
  {journal} {\bibinfo  {journal} {Nature Nanotechnology}\ }\textbf {\bibinfo
  {volume} {8}},\ \bibinfo {pages} {321} (\bibinfo {year} {2013})}\BibitemShut
  {NoStop}%
\bibitem [{\citenamefont {Arnold}\ \emph {et~al.}(1998)\citenamefont {Arnold},
  \citenamefont {Wilson},\ and\ \citenamefont {Boshier}}]{Arnold1998}%
  \BibitemOpen
  \bibfield  {author} {\bibinfo {author} {\bibfnamefont {A.~S.}\ \bibnamefont
  {Arnold}}, \bibinfo {author} {\bibfnamefont {J.~S.}\ \bibnamefont {Wilson}},
  \ and\ \bibinfo {author} {\bibfnamefont {M.~G.}\ \bibnamefont {Boshier}},\
  }\href {\doibase 10.1063/1.1148756} {\bibfield  {journal} {\bibinfo
  {journal} {Review of Scientific Instruments}\ }\textbf {\bibinfo {volume}
  {69}},\ \bibinfo {pages} {1236} (\bibinfo {year} {1998})}\BibitemShut
  {NoStop}%
\bibitem [{\citenamefont {Grove}\ \emph {et~al.}(1995)\citenamefont {Grove},
  \citenamefont {Sanchez-Villicana}, \citenamefont {Duncan}, \citenamefont
  {Maleki},\ and\ \citenamefont {Gould}}]{Grove1995}%
  \BibitemOpen
  \bibfield  {author} {\bibinfo {author} {\bibfnamefont {T.~T.}\ \bibnamefont
  {Grove}}, \bibinfo {author} {\bibfnamefont {V.}~\bibnamefont
  {Sanchez-Villicana}}, \bibinfo {author} {\bibfnamefont {B.~C.}\ \bibnamefont
  {Duncan}}, \bibinfo {author} {\bibfnamefont {S.}~\bibnamefont {Maleki}}, \
  and\ \bibinfo {author} {\bibfnamefont {P.~L.}\ \bibnamefont {Gould}},\ }\href
  {\doibase 10.1088/0031-8949/52/3/008} {\bibfield  {journal} {\bibinfo
  {journal} {Physica Scripta}\ }\textbf {\bibinfo {volume} {52}},\ \bibinfo
  {pages} {271} (\bibinfo {year} {1995})}\BibitemShut {NoStop}%
\bibitem [{\citenamefont {Rushton}\ \emph {et~al.}(2014)\citenamefont
  {Rushton}, \citenamefont {Aldous},\ and\ \citenamefont
  {Himsworth}}]{rushton}%
  \BibitemOpen
  \bibfield  {author} {\bibinfo {author} {\bibfnamefont {J.~A.}\ \bibnamefont
  {Rushton}}, \bibinfo {author} {\bibfnamefont {M.}~\bibnamefont {Aldous}}, \
  and\ \bibinfo {author} {\bibfnamefont {M.~D.}\ \bibnamefont {Himsworth}},\
  }\href {\doibase 10.1063/1.4904066} {\bibfield  {journal} {\bibinfo
  {journal} {Review of Scientific Instruments}\ }\textbf {\bibinfo {volume}
  {85}},\ \bibinfo {pages} {121501} (\bibinfo {year} {2014})}\BibitemShut
  {NoStop}%
\bibitem [{\citenamefont {Wright}\ and\ \citenamefont
  {Gianchandani}(2007)}]{wright}%
  \BibitemOpen
  \bibfield  {author} {\bibinfo {author} {\bibfnamefont {S.~A.}\ \bibnamefont
  {Wright}}\ and\ \bibinfo {author} {\bibfnamefont {Y.~B.}\ \bibnamefont
  {Gianchandani}},\ }\href {\doibase 10.1116/1.2782510} {\bibfield  {journal}
  {\bibinfo  {journal} {Journal of Vacuum Science \& Technology B:
  Microelectronics and Nanometer Structures Processing, Measurement, and
  Phenomena}\ }\textbf {\bibinfo {volume} {25}},\ \bibinfo {pages} {1711}
  (\bibinfo {year} {2007})}\BibitemShut {NoStop}%
\bibitem [{\citenamefont {Grzebyk}\ \emph {et~al.}(2014)\citenamefont
  {Grzebyk}, \citenamefont {G{\'o}recka-Drzazga},\ and\ \citenamefont
  {Dziuban}}]{grzebyk}%
  \BibitemOpen
  \bibfield  {author} {\bibinfo {author} {\bibfnamefont {T.}~\bibnamefont
  {Grzebyk}}, \bibinfo {author} {\bibfnamefont {A.}~\bibnamefont
  {G{\'o}recka-Drzazga}}, \ and\ \bibinfo {author} {\bibfnamefont {J.~A.}\
  \bibnamefont {Dziuban}},\ }\href {\doibase
  https://doi.org/10.1016/j.sna.2014.01.011} {\bibfield  {journal} {\bibinfo
  {journal} {Sensors and Actuators A: Physical}\ }\textbf {\bibinfo {volume}
  {208}},\ \bibinfo {pages} {113} (\bibinfo {year} {2014})}\BibitemShut
  {NoStop}%
\end{thebibliography}%

\end{document}